# Studying the Effects of Deep Brain Stimulation and Medication on the Dynamics of STN-LFP Signals for Human Behavior Analysis

Hosein M. Golshan, Adam O. Hebb, Joshua Nedrud, and Mohammad H. Mahoor

*Abstract*— This paper presents the results of our recent work on studying the effects of deep brain stimulation (DBS) and medication on the dynamics of brain local field potential (LFP) signals used for behavior analysis of patients with Parkinson's disease (PD). DBS is a technique used to alleviate the severe symptoms of PD when pharmacotherapy is not very effective. Behavior recognition from the LFP signals recorded from the subthalamic nucleus (STN) has application in developing closed-loop DBS systems, where the stimulation pulse is adaptively generated according to subjects' performing behavior. Most of the existing studies on behavior recognition that use STN-LFPs are based on the DBS being "off". This paper discovers how the performance and accuracy of automated behavior recognition from the LFP signals are affected under different paradigms of stimulation on/off. We first study the notion of beta power suppression in LFP signals under different scenarios (stimulation on/off and medication on/off). Afterwards, we explore the accuracy of support vector machines in predicting human actions ("button press" and "reach") using the spectrogram of STN-LFP signals. Our experiments on the recorded LFP signals of three subjects confirm that the beta power is suppressed significantly when the patients take medication (*p*-value<0.002) or stimulation (*p*-value<0.0003). The results also show that we can classify different behaviors with a reasonable accuracy of 85% even when the high-amplitude stimulation is applied.

## I. INTRODUCTION

Parkinson's disease (PD) is a neurodegenerative condition diagnosed on the basis of clinical history and motor signs of tremor, rigidity and bradykinesia (slow movements). PD incidence increases with advancing age and peaks among people in their 60s and 70s. It is known that in patients with PD, the loss of nigral dopaminergic input to the striatum leads to motor disorders as a result of augmented beta frequency (10–30 Hz) oscillatory power [1]. Lack of dopamine in the basal ganglia increases the synchronized oscillatory activity in the beta frequency of subthalamic nucleus local field potential (STN-LFP) recordings, which correlates with worsening symptoms of PD [2].

Deep brain stimulation (DBS) surgery is often used for patients with severe signs of PD, specifically when pharmacotherapy no longer mitigates the patients' symptoms effectively. The surgical procedure consists of implanting DBS leads in the globus pallidum (GPi) or STN of the brain for neuro-stimulation by a high-frequency (~130-185 Hz) electrical pulse [1].

Although DBS is effective in alleviating the motor symptoms of PD, it may lead to side effects such as impaired cognition, speech, and balance disruptions [3,4]. An adaptive (closed-loop) DBS system that modulate the stimulation pulse based on the brain feedbacks and patients' current goal may reduce these side effects, while providing maximum therapeutic benefit for PD motor symptoms [5,6].

Recognition of human behavior through brain signals is a necessary step to design such a closed-loop system. Santaniello *et al.*, [7] proposed a closed-loop DBS system to adjust the stimulation amplitude using LFP signals acquired from ventral intermediate nucleus (VIM) of the thalamus. Loukas and Brown [8] proposed an algorithm to predict self-paced hand-movements based on the oscillatory nature of the STN-LFPs. Time-frequency analysis of the LFP signals in the beta frequency range has been used in different studies [9-12] to perform the behavior recognition task.

LFPs recorded from the STN are robust control signals to indicate a change in patient's state [11,13]. Moreover, STN-LFPs correlate with symptoms of PD, levodopa medication level, behavior, and neuro-stimulation intensity [1,14]. PD symptoms may vary depending on patient's level of attention and behavior, which change the characteristics of STN-LFP activity consequently. Therefore, the chronical variability of recorded LFPs must be investigated to design a robust closed-loop DBS system.

In this paper, we chronically recorded STN-LFP from three PD subjects in the medication on/off and stimulation on/off states while the subjects performed different motor tasks such as pressing a button and reaching a target. As a consequence, it helps explore the dynamics (i.e., change of characteristics) of LFP signals with different therapeutic conditions within a long time window. In contrast to our previous studies [9-12], where the LFP signals were recorded in the operating room, we use a research-grade implantable neuro-stimulator (INS) to bilaterally record LFPs outside the operating room. Moreover, as opposed to other related works [9-12] that utilize the LFP signals with stimulation "off" condition, this study aims to evaluate the capability of human behavior recognition when stimulation is "on".

* This research is partially supported by the Knoebel Institute for Healthy Aging (KIHA) at University of Denver, Denver, CO, USA.

Hosein M. Golshan is a Ph.D. student in the ECE department, University of Denver, Denver, CO, USA. (email: hosein.golshanmojdehi@du.edu)

Adam O. Hebb is a neurosurgeon at Kaiser Hospital, and a research scholar in the ECE department, University of Denver, Denver, CO, USA. (email: adam.hebb@aoh.md)

Joshua Nedrud is an investigator scientist at Colorado Neurological Institute (CNI), Englewood, CO, USA.(email: jnedrud@thecni.org)

Mohammad H. Mahoor is an Associate Professor in the ECE department, University of Denver, Denver, CO, USA. (email: mohammad.mahoor@du.edu)

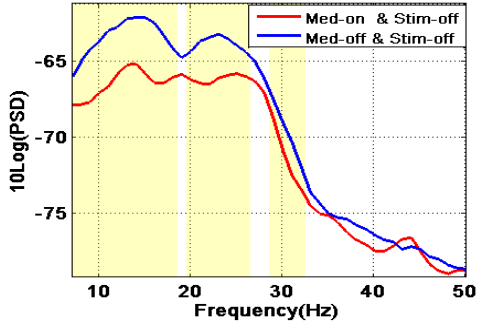 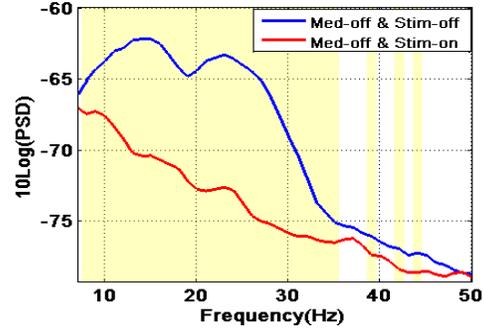

Fig. 2. Comparison the effect of medication (left red) and stimulation (right red) on the beta power of LFP signals (blue) recorded from a PD subject performing "button press" trials. The graphs are obtained by averaging the PSD of 60 recordings. The highlighted yellow area shows the frequency range where the difference between two overlaid graphs (blue and red) is significant (i.e., $p$-value $< 0.05$).

First, the power spectrum density (PSD) of the recorded LFPs is analyzed to compare the effect of the medication and stimulation pulse on the increased beta power associated with PD. Then, we use the time-frequency representation of the raw LFP data to design a feature space based on which the behavior classification task can be done with both stimulation "off" and "on" conditions. We use support vector machine (SVM) classifiers to classify "button press" and "reach" trials with different compositions of the data, including stimulation "off", "on", and "off-on combined". This provides valuable insight into the reliability of the classification of LFP signals in different stimulation conditions.

The rest of this paper is organized as follows: Section II elaborates on the data recording sessions. The methodology is presented in Section III. The quantitative results are given in Section IV. Finally, Section V comprises discussion and some remarks.

## II. MATERIALS AND DATA RECORDING

Three PD subjects with an implanted DBS lead (Medtronic 3389, Minneapolis, MN, USA) in each STN and subcutaneously implanted INS (Activa PC+S, Medtronic Inc.,) participated in this study. All participants provided informed consent in a manner approved by the HealthOne Institutional Review Board.

All subjects underwent postoperative data recording sessions. During the behavioral recording, two bipolar referenced LFPs, one from each hemisphere, were amplified and digitized ($F_s \sim 422$Hz) by the INS. In the experiments, the bipolar pair of channels containing the most prominent peak in beta frequency range was selected for recordings. Moreover, the INS generates a stimulation pulse with amplitude ~2.5v, frequency ~140Hz, and pulse width ~60μs.

Recordings were performed at 12 or 24 months after DBS lead implantation surgery. The first recording session was performed while the subjects refrained from taking their levodopa medication for at least 12 hours. The second session was performed when the subjects were regularly consuming their prescribed medication dosage. On average, behavioral tasks included 60 repetitions of cued "button press" and "target reaching" trials performed by left or right hands, under stimulation "off" and "on" conditions.

## III. METHODOLOGY

In this section, first, the effect of the medication and stimulation pulse on the beta power will be presented. Then, we explain the feature extraction and classification approach.

### A. Analysis of Power Spectrum Density (PSD)

To compare the therapeutic effect of medication and stimulation on the increased beta power associated with PD, the power spectrum density (PSD) of the recorded LFPs are calculated using a custom Welch's method with a Hanning window of length nfft = $F_s$. This provides a frequency resolution ($F_s$ / nfft) of 1Hz. The PSD of an LFP signal $x(t)$ is given by its corresponding discrete Fourier transform $X(f)$ coefficients as follows [15]:

$$PSD(f) = \frac{1}{F_s \times S}|X(f)|^2, \quad S = \sum_{i=1}^{N} \omega_i^2. \quad (1)$$

where, $S$ is the scaling factor defined as the sum of squared weights ($\omega$) of the employed window (Hanning window).

The impact of stimulation on the PSD of an exemplary LFP signal is given in Fig. 1. As shown, there is a significant peak about the stimulation frequency ($f \sim 140$Hz), which has been propagated into its neighbor frequency components $f \sim (100-180)$Hz. However, the low frequency range is not considerably affected by the stimulation pulse, except for the expected therapeutic decrease on the beta power.

To evaluate the effect of medication and stimulation pulse on the beta power of PD patients, we use the one-way analysis of variance (ANOVA) statistical test. In this way, the significance of difference between the beta power for each case, (medication off vs medication on) and (stimulation off vs stimulation on), can be measured (see Fig. 2). The quantitative results are presented in Section IV.

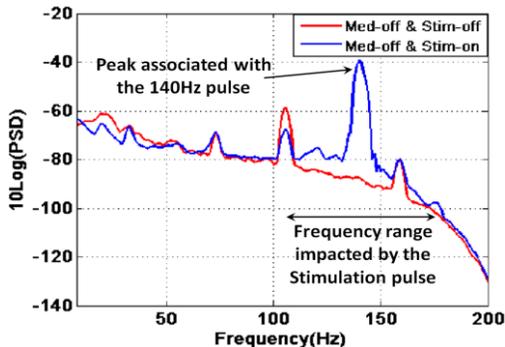

Fig. 1. Comparison of PSDs between stimulation "off" (red) and "on" (blue) cases. As seen, the artifact imposed by the stimulation pulse mainly impacts the high frequency range $f \sim (100-180)$Hz. At the low frequency range, the beta power decreases as a result of the stimulation.

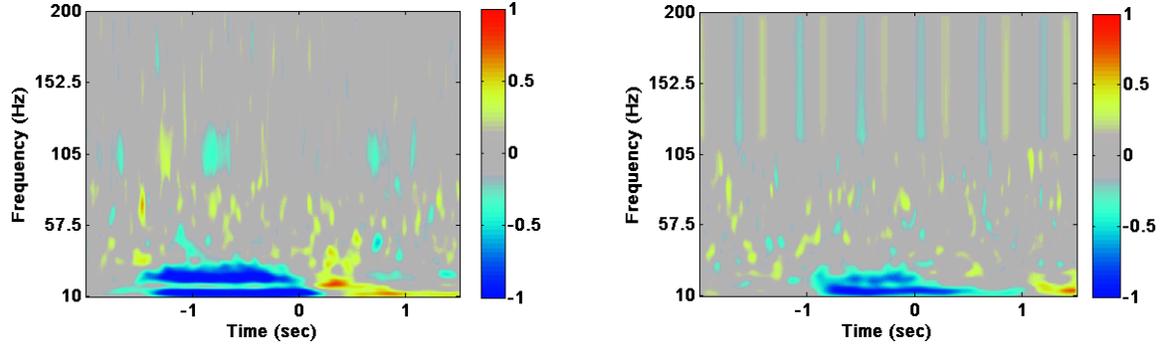

Fig. 3. The average amplitude of wavelet coefficients calculated for LFP signals of 60 "reach" trials. The results are shown for a time interval (-2,1.5)sec around each onset centered on 0sec. Left and right respectively shows the spectrogram obtained under stimulation "off" and "on" conditions. Consider the similarity of patterns at the beta frequency (both figures) and the artifact generated by the stimulation at $f$~(100-200)Hz (right).

### B. Classification under Stimulation off and on conditions

The time-frequency representation of the raw LFP signals is a reliable feature space to distinguish different human behaviors [9-12]. Hence, we use continuous wavelet transform (CWT) of the recorded LFPs to generate a feature space based on which the behavior classification task can be done. A complex Morlet wavelet function ($\psi(t) = \exp(-t^2/f_b)\exp(j2\pi f_c t)/\sqrt{\pi f_b}$; $f_c$: central frequency, $f_b$: bandwidth) is employed to obtain the time-frequency representation of the data. Fig. 3 shows the average amplitude of the CWT coefficients of all 60 trials recorded form one of the subjects and "reach" task. As seen, except for some artifacts appeared at high-frequency components (i.e., about the stimulation frequency $f$~140Hz), the spectrograms given for both stimulation "off" and "on" cases follow the same patterns. It suggests that, regardless of the stimulation status, the behavior classification task can be done properly as long as the low frequency components are used.

In this paper, we employ the SVM classifier [16] for the behavior classification purpose. The SVM classifier tends to separate dataset ($D = \{\mathbf{x}_i, y_i\}_{i=1}^N$, $\mathbf{x}_i \in R^d$, $y_i \in \{-1, +1\}$) by learning an optimal hyperplane $\langle \mathbf{w}, \mathbf{x} \rangle + b = 0$ between classes such that the margin between them becomes maximum; where, $\mathbf{w}$ and b are the parameters of the hyperplane, $\langle \cdot, \cdot \rangle$ denotes inner product operation, $\mathbf{x}_i$ and $y_i$ respectively represent the $i^{th}$ sample and its corresponding label. In general, the classifier is obtained by minimizing the following equation with respect to $\mathbf{w}$:

$$\min \frac{1}{2}\|\mathbf{w}\|^2 + C\left(\sum_{i=1}^N \xi_i\right) \quad s.t. \quad y_i(\langle \mathbf{w}, \mathbf{x}_i \rangle + b) \geq 1 - \xi_i. \quad (2)$$

$$\text{Test label} = \text{sgn}\left(\sum_{i=1}^N \alpha_i y_i k(\mathbf{x}_i, \mathbf{x}_{\text{test}}) + b\right)$$

where, $\xi_i \geq 0$ and $C \geq 0$ are respectively the slack and penalty variables. $\alpha_i$ is the Lagrangian multiplier, $\text{sgn}(\cdot)$ is the sign function, and $k(\cdot, \cdot)$ is a kernel function to project the data on a new space where the linear classification becomes possible.

### IV. EXPERIMENTS AND RESULTS

In this section, first, the performance of the proposed wavelet-based feature space in classifying two behavioral tasks ("button press" and "reach") under stimulation "off", "on", and "off-on combined" conditions is examined. To this end, the beta frequency components of the raw LFPs acquired from three PD subjects (Section II) are calculated within an interval (-1,1)sec around each task onset. To keep the computational burden low, principal component analysis (PCA) is applied on the calculated feature vectors before exposing them to the classifier [11]. In our experiments, 95% of the eigenvalues corresponding to the maximum variance direction is kept with the PCA calculations.

The quantitative results given by the SVM classifier with RBF kernel ($k(\mathbf{x}, \mathbf{y}) = exp(\gamma\|\mathbf{x}-\mathbf{y}\|^2)$) and linear kernel ($k(\mathbf{x}, \mathbf{y}) = \mathbf{x}^T\mathbf{y}+c$)), as well as KNN classifier are tabulated in Table I. In each case, the classification accuracy (truly predicted labels /ground-truth), precision (true positive / (true positive + false positive)), recall (true positive / (true positive + false negative)), and AUC (area under the receiver operating characteristic curve) are reported. Note that the parameters of the classifiers are experimentally set so as to achieve the best performance, i.e., linear-SVM C = 100; RBF-SVM C = 100 and $\gamma$ = 0.001; KNN K = 3.

As Table I shows, the RBF-SVM classifier returns the best results for the quantitative measures. The results obtained under stimulation "on" case are comparable to the results of stimulation "off" case, with a margin of 2%. It shows that, regardless of the artifact imposed by the high-amplitude stimulation pulse, the behavior classification can be done using the informative beta band patterns. Moreover, the classification performance approximately remains unchanged when then stimulation "off-on combined" case is compared to stimulation "off" case (the stimulation "on" data gives very competitive results). For instance, the classification accuracy is 87%, 85%, and 87% under stimulation "off", "on", and "off-on combined" conditions.

TABLE I. COMPARISON OF CLASSIFICATION PERFORMANCE (%) IN CLASSIFYING "BUTTON PRESS" AND "REACH" TRIALS WITH DIFFERENT COMPOSITIONS OF LFP DATA, INCLUDING STIMULATION "OFF", "ON", AND "OFF-ON COMBINED". THE BEST RESULTS ARE HIGHLIGHTED IN EACH CASE.

|  | SVM-Linear | | | | SVM-RBF | | | | KNN (K=3) | | | |
| --- | --- | --- | --- | --- | --- | --- | --- | --- | --- | --- | --- | --- |
|  | Accuracy | Precision | Recall | AUC | Accuracy | Precision | Recall | AUC | Accuracy | Precision | Recall | AUC |
| Stim-off | 84 | 84 | 83 | 91 | **87** | **88** | **88** | **94** | 85 | 87 | 82 | 84 |
| Stim-on | 81 | 82 | 80 | 89 | **85** | **85** | **84** | **92** | 81 | 84 | 79 | 81 |
| Stim-off & on | 84 | 84 | 83 | 90 | **87** | **87** | **88** | **94** | 83 | 84 | 82 | 84 |

In addition, the efficiency of medication and stimulation on the beta power of PD patients is separately compared here. Based on our analysis on the PSD of the raw LFP data calculated within an interval (-0.5,0.5)sec around each task onset, the increased beta power associated with PD is significantly reduced under medication ($p$-value $< 0.002$) and stimulation pulse ($p$-value $< 0.0003$). Note that the $p$-values are calculated based on the average $p$-values of a frequency interval around the peak of PSDs (i.e., $\sim f_{peak} \pm 5$ Hz) across all subjects and trials. For instance, the frequency range of $f \sim (10\text{-}20)$Hz is used for Fig. 2.

Considering the experimental results, the beta power decrease is more significant in almost all cases when the stimulation pulse is "on". As opposed to the medication "on" cases, the $p$-values remains below 0.05 approximately throughout the beta frequency range when the stimulation pulse is applied (see Fig. 4). This fact can show the prominence of the stimulation pulse over the medication therapy in tackling the PD symptoms.

## IV. CONCLUSION AND DISCUSSION

In this paper, the effect of the stimulation pulse and medication on the human behavior classification and beta power of Parkinson's disease (PD) patients was evaluated. A feature space based on the time-frequency representation of the acquired brain subthalamic nucleus local field potential (STN-LFP) signals was developed. Our analysis showed that the beta frequency components of LFPs are capable of detecting different human activities even when the high-amplitude deep brain stimulation (DBS) pulse is applied.

Different experiments were carried out on the LFP signals acquired from three PD subjects to classify "button press" and "reach" trials. The performance of behavior classification was evaluated under stimulation "off", "on", and "off-on combined" conditions. The results showed that, regardless of the stimulation status, the behavior classification capability remains almost unchanged when the beta frequency components of the proposed feature space are used. As a result, there is no need to remove the high-frequency artifacts imposed by the stimulation pulse, which essentially requires more computational power.

Furthermore, the effect of stimulation and medication on the beta power of LFP signal was separately investigated. The results showed that the stimulation artifact mainly impacts the frequency range around the stimulation frequency. Also, the increased beta power associated with PD is suppressed significantly when the patients take medication or receive therapeutic stimulation.

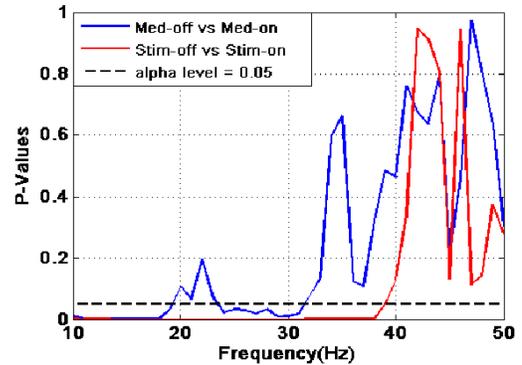

Fig. 4. Variations of $p$-values with respect to the frequency range for "reach" trial. (Blue) and (red) graphs respectively depict the significance of difference (between medication "off" & "on" while stimulation is "off") and (between stimulation "off" & "on" while medication is "off"). As seen, $p$-values remain less than 0.05 for the entire beta range $f \sim (10\text{-}30)$Hz when the stimulation pulse is on (red).